# Correlated conformation and charge transport in multiwall carbon nanotube - conducting polymer nanocomposites

Paramita Kar Choudhury[1], S. Ramaprabhu[2], K. P. Ramesh[1], and Reghu Menon[1]

[1]*Department of Physics, Indian Institute of Science, Bangalore 560012, India*
[2]*Department of Physics, Indian Institute of Technology Madras, Chennai-600036, India*

The strikingly different charge transport behaviors in nanocomposites of multiwall carbon nanotubes (MWNTs) and conducting polymer polyethylene dioxythiophene – polystyrene sulfonic acid (PEDOT-PSS) at low temperatures are explained by probing their conformational properties using small angle X-ray scattering (SAXS). The SAXS studies indicate assembly of elongated PEDOT-PSS globules on the walls of nanotubes, coating them partially thereby limiting the interaction between the nanotubes in the polymer matrix. This results in a charge transport governed mainly by small polarons in the conducting polymer despite the presence of metallic MWNTs. At T > 4 K, hopping of the charge carriers following 1D-VRH is evident which also gives rise to a positive magnetoresistance (MR) with an enhanced localization length (~ 5 nm) due to the presence of MWNTs. However, at T < 4 K, the observation of an unconventional positive temperature coefficient of resistivity (TCR) is attributed to small polaron tunnelling. The exceptionally large negative MR observed in this temperature regime is conjectured to be due to the presence of quasi-1D MWNTs that can aid in lowering the tunnelling barrier across the nanotube – polymer boundary resulting in large delocalization.

Email: s_paramita@physics.iisc.ernet.in

## 1. Introduction

Since the high processibility and unique optoelectronic properties of conducting polymers can be complimented with striking electronic, mechanical and thermal properties of carbon nanotubes, there has been a momentous effort in harvesting the goodness of both these systems: in the form of nanocomposites. Both conducting polymers and carbon nanotubes (CNTs) are one-dimensional (1D) systems that consist of delocalized π-electrons. The main difference between these systems, however, is the fact that flexibility in conducting polymers can be controlled by chemical modifications while CNTs are rigid. These conformations and the extent of delocalization of $\pi$-electrons in these quasi-one dimensional systems play significant roles in the electrical properties of these systems.

The contrast in electron density between semi-flexible conjugated polymer chains and rigid-rod carbon nanotubes has been imaged using SEM and TEM; more precise light and X-ray scattering experiments have been used to study the morphology of nanotube suspensions and nanocomposites in polymers [1,2]. But the previous studies have not converged to a consistent picture yet. Zhou et al have reported a rigid rod structure of nanotube suspension in $D_2O$ while observation of a rope-like disordered fractal objects have been reported by Schaefer et al., especially when polyelectrolytes are used as dispersion aids for the suspension [1,2]. Moreover, small angle neutron scattering (SANS) study on surfactant aided aqueous dispersion of SWNTs did not show any significant contribution of nanotubes in the formation of micellar structure of surfactant molecules in water [3].

In such complex systems, therefore, it is indeed a challenge to separate out the various contributions to charge transport and other physical properties; however the dispersibility of the tubes and inter-tube barriers have been identified as the key ingredients in limiting the bulk transport. In well-processed nanocomposites, the typical percolation threshold is 0.1 – 1 wt. % of CNT, with conductivity $10^{-2}$ – 1 S/cm [4]. Although a wide variation in the temperature dependence of conductivity has been observed in these systems, the conductivity value at room temperature usually decreases by several orders of magnitude at 4.2 K. There have been attempts to prepare the CNT composites in conducting polymer matrices so that the contributions from both can weaken the temperature dependence of conductivity and to make the system more metallic [5,6]. In these cases either polyaniline or polypyrrole is used as the conducting matrix, and the temperature dependence of conductivity show significant changes only at large volume fractions (~ 10 %) of CNTs [5,7]. This is because in CNT composites with conducting polymers, the sharp percolation threshold for the onset of increase in conductivity is not observed, which is obvious since the conducting matrix also contributes to the overall charge conduction. Also, the mixing of conduction mechanisms via the CNTs and conducting polymer makes it hard to discern the exact ones, unlike in case of CNT-insulating polymer composites. This complicated scenario is even true in various types of CNT samples due to the wide variation in impurities, defects, size, packing and ordering of the nanotubes. Apart from these intrinsic traits in CNTs, the details of the inter-tube barriers come into play in CNT-polymer composites, and this affects the localization and interaction contributions to charge transport. Although models like fluctuation induced tunnelling (FIT), hopping of localized charge carriers (variable range hopping - VRH) or a combination of both have been widely used to analyse the data in CNT-polymer systems [4,8], as the complexity of the system increases, the intriguing results cannot be easily explained with the help of conventional models. In such cases understanding the conformation and organization of the system can throw light on the possible transport mechanisms. However, there are very few experimental studies in conjugated polymer-CNT composites exploring how the conformational features are reflected in their unique electronic transport.

In this work, conformation of multiwall carbon nanotube (MWNT) suspensions in aqueous solution of conducting polymer polyethylene dioxythiophene – polystyrene sulfonic acid (PEDOT-PSS) is studied using small angle X-ray scattering (SAXS) technique. Small angle X-ray scattering (SAXS) studies in polymeric systems have already shown that the local nanoscale morphology at various length scales can be probed to show the correlation among conformation and assembly of chains [9-11]. Our results show rigid-rod characteristics of nanotubes held in a meshwork in the polymer solution; the polymer globules close to the CNTs are extended and tend to aggregate onto the nanotube walls partially "coating" them. The memory of the conformation in the suspension state is retained in the solid nanocomposite films.

Furthermore, the interplay of conduction mechanisms of the CNTs embedded in the PEDOT-PSS matrix is investigated. Since PEDOT-PSS is widely used as transparent electrode in polymer and organic devices, the low-temperature conductivity and magnetoresistance measurements have been carried out in MWNT-PEDOT-PSS composite films. As said earlier, the SAXS results indicate that the PEDOT-PSS globules tend to adhere to the walls of CNTs. This could weaken the inter-tube transport especially in the lower volume fraction range. In case of higher volume fractions the CNTs tend to form aggregates that make the system highly inhomogeneous, and the intrinsic mesoscopic properties of individual tubes are usually lost due to presence of intertube screening of too many nanotubes. Due to these factors the concentration of nanotubes in PEDOT-PSS in this study is limited to 0.03 - 3 %. The lower cut-off for concentration is verified to be above pecolation threshold. Because of the restricted CNT concentration, individual nanotubes as well as the conducting polymer matrix are expected to contribute substantially to the overall charge transport and very interesting indications of the contribution to transport mediated via the CNTs are observed at low temperatures.

## 2. Experiment

The average diameter of MWNTs used is ~ 40 nm extending up to a length of a few hundred nanometers, and the samples were prepared by CVD techniques [12]. Short PEDOT chains complexed onto PSS template (PEDOT-PSS, from Baytron-PTM) was used as the matrix of the nanocomposite. For the SAXS measurements, aqueous solution of 1.1 wt. % PEDOT + PSS was filtered using a 0.2 mm PTFE membrane filter to remove traces of macroscopic aggregates. 1 wt. % of acid washed and dried CNTs were dispersed in triple distilled water (TDA) by unltrasonication for 15 minutes, and the suspension was named CNTW. The aqueous suspensions of CNTs were then added to PEDOT-PSS in water and ultrasonicated to prepare dispersions of 0.03, 0.3 and 1 wt. % CNT with respect to PEDOT-PSS and the samples were named CNTP - (0.03, 0.3, and 1 %). The SAXS studies were limited to samples containing less than 1 % of CNT, so that they are within the dilute solution limit so as to restrict the aggregation of CNTs. Aqueous solution of the pristine polymer PEDOT-PSS was also studied as reference. All the samples were filled in Mark glass capillary of 2 mm diameter and sealed.

SAXS measurements were carried out using Bruker Nanostar equipped with a rotating anode source and three-pinhole collimation. A position sensitive 2D detector with 100 μm resolution was used to record the scattered intensity. The scattered intensity *I(q)* is plotted as a function of the momentum transfer vector $q = 4\pi \sin\theta / \lambda$, where $\lambda$ is the wavelength of the X-rays (Cu-kα radiation, 1.54 Å), and $\theta$ is half the scattering angle. The q-range is 0.008 Å$^{-1}$ < q < 0.3 Å$^{-1}$. The raw data was normalized for transmission coefficient, capillary width and exposure time; the incoherent scatterings due to solvent were subtracted in the data analysis.

For the conductivity and magnetoresistance measurements on nanocomposites, 0.03, 0.3 and 3 wt. % of CNT suspensions in water were dispersed in aqueous PEDOT/PSS solution by ultrasonication for 30 minutes. Free standing nanocomposite films of thickness 15 – 20 μm were obtained from the CNT-PEDOT/PSS solution by drop casting on glass substrate and eventual evaporation at 60-70 °C. The TEM image in Fig. 1 shows the network of MWNTs in the polymer matrix.

The electrical measurements were performed on the 0.03, 0.3 and 3 % nanocomposites with standard four-probe dc method in a Janis variable temperature cryogenic system equipped with an 11 T superconducting magnet. The magnetic field was applied parallel to the network plane. The current used in low temperature transport measurements are below 1 μA, and the heat dissipation is typically less than 50 nW. The temperature was stable to within 20 mK during the field sweep. Standard four-probe measurements were also carried out on pressed pellets of MWNTs and the data was used as reference to the nanocomposite data.

## 3. Results and discussion

I vs. q profiles of 0.03, 0.3, and 1 wt. % CNTs dispersed in aq. PEDOT-PSS (CNTP) are displayed in Fig. 2. Inset shows I vs. q plot for 1 % CNTW suspension. The slope ~ -1 for a decade of $q$ range 0.023 Å$^{-1}$ < $q$ < 0.2 Å$^{-1}$ indicates rigid-rod like structure of MWNTs in agreement with earlier reports [2]. The $q$-range of the data is not sufficient to probe the 'microns-long' length scales of a nanotube since the persistence length is presumably close to the length of a nanotube; and this slope represents the partial rigid structure present in the aqueous suspension. However, for $q$ < 0.023 Å$^{-1}$ the slope sharply changes to ~ -2.6. In this larger length scale, the resultant scattering from more than a single nanotube, and their intersections can be observed. The slope -2.6 indicates the presence of an interconnected loose 3-D meshwork, also usually reported for semi-rigid polymers. The scattering profile of the nanocomposite might resemble more to that of the semi-rigid polymer when the size of the meshwork matches with the persistence length of the polymer chains.

For the three CNTP scattering profiles, the slope in the range 0.023 Å$^{-1}$ < $q$ < 0.2 Å$^{-1}$ shows a deviation from -1. This is due to the compact-coil globular structure of PEDOT-PSS in aqueous solution [13] in addition to the rod-like CNTs. Since the length scales of the polymer is much smaller (radius of gyration: $R_g$ ~ 3 - 4 nm) with respect to the cross-section radius of CNTs (~ 10 - 20 nm), the system can be viewed as a sea of polymer globules with very few weakly interacting nanotubes. The slope in the same range for different nanocomposites shifts towards smaller values as the nanotube concentration decreases. This is expected since the number of rigid scatterers decreases, and the contribution to the net intensity reduces with respect to that from flexible polymers. The slope ~ -2.8 for $q < 0.023$ Å$^{-1}$, similar to that for CNTW, shows that the network of CNTs still holds among the surrounding PEDOT-PSS globular structures in the nanocomposite systems.

In order to understand the modification in the conformation of PEDOT-PSS in the nanocomposite, the data of 1 % CNTP data is studied as a typical case. The CNTW data is subtracted from this CNTP data and the resultant CNTP – CNTW data, which represents the scattering profile of the polymer in the nanocomposite, is shown in Fig. 3. The data for pristine PEDOT-PSS in water (1 % by weight) is shown in the inset for comparison. It is interesting to see that the conformation of the polymer chains in the nanocomposite is quite different from that in the aqueous solution. The fractional slope of -2.68, reported as signature of mass-fractal in aq. PEDOT-PSS [13], is also present in the nanocomposite, but at lower $q$-range. The Guinier region that was observed for aq. PEDOT-PSS is not seen in the present case; and the scales apparently seem to be shifted towards larger lengths.

The most striking difference between the two cases is the presence of a plateau for the $q$-range 0.02 Å$^{-1}$ < $q$ < 0.2 Å$^{-1}$ for PEDOT-PSS in nanocomposite which is not present in its aqueous solution. The slope ~ -0.5 is not indicative of any particular structure; also the sharp upturn to the slope -2.8 at q < 0.022 Å$^{-1}$ signifies presence of ordering or self-assembly at larger length scales. To investigate this further, the pair distribution function (PDF) analysis can be of immense help since the features in reciprocal space ought to reflect that in the real space; also providing quantitative values of structural parameter like $R_g$.

The pair distribution function *p(r)* is calculated by inverse Fourier transform of the scattered intensity I(q), using the algorithm GNOM [14] :

$$p(r) = \frac{1}{2\pi^2} \int_{q_{min}}^{q_{max}} I(q) qr \sin(qr) dq \qquad (1)$$

Depending on the range and nature of ordering in the system, *p(r)* can have many peaks; the first maximum occurs for the most probable distance in the scattering chain, which tends to zero at $r_{max}$, the maximum intra-chain distance. We can thus compute the average size of chains *<Rg>* from the normalized second moment of *p(r)*, i.e.:

$$<R_g>^2 = \frac{\int_0^{r_{max}} r^2 p(r) dr}{2 \int_0^{r_{max}} p(r) dr} \qquad (2)$$

The data for P(r) vs. r for PEDOT-PSS in nanocomposite is shown in Fig. 4, and the data is compared to that for aqueous PEDOT-PSS. The PDF profiles in Fig. 4, clearly shows a marked difference in conformation and organization of the polymer chains in presence of MWNTs. The first peak representing the intra-chain correlation is prominently present up to 115 Å in the aqueous solution, while it is diminished to about 25 Å in the nanocomposite. This suggests the possibility that the compact coiled structure of those adhered to the nanotube walls is getting modified to form more extended structures in presence of CNTs. Also, the position of maximum value of r for the first peak $r_1 = (r_{max})/2$, that represent a spherical particle. It is known that the ratio $r_1/r_{max}$ decreases further with increasing elongation of the particle. In case of PEDOT-PSS in nanocomposite, this ratio (0.07) is very small, indicating the presence of elongated structures. The most interesting feature in case of PEDOT-PSS in nanocomposite, however, is the pronounced second peak that represents inter-chain correlations; the peak height being more than twice that of the first peak and a large area under the curve. Such inter-chain correlation has been observed in aqueous PEDOT-PSS, but not to such a large extent with respect to the intra-chain correlation.

The pronounced second peak gives evidence for strong correlation among the extended globules around the nanotube walls. The detailed electron microscopy studies have shown that such large-scale organization is possible in these systems [15]. The data analysis indicates that the correlation among these structures can extend up to 350 Å ~ 35 nm, which incidentally is the diameter of the nanotubes. However, the shoulder-like features observed on either side of the second peak indicate an underlying shallow feature that persists to a larger length scale. Although the origin for this not very clear, but it may be interpreted as some sort of macroscopic scale assembly of the extended globules. The PDF analysis for PEDOT-PSS in nanocomposites gives an average value of $R_g$ among the extended (close to the nanowalls) and globular (away from the nanowalls) structures. This average estimate of $R_g \sim 12.5$ (± 0.5) Å is three times larger than that observed in case of aqueous solution (~ 3.8 Å). Thus, the presence of CNTs modifies the conformation and assembly of PEDOT-PSS globules, which can be used to control the assembly of chains in the liquid state and the same information can be carried forward to the solid films cast from the solution.

It is noteworthy that the CNTs after preparation and acid cleaning are left with groups like -COOH on their walls that enable them to get dispersed evenly in water without the use of any surfactants. The PEDOT-PSS complex, on the other hand, has –$SO_3H$ side groups, which can form hydrogen bonds with -COOH groups on nanotube wall [16-19]. The rigid-rod character of nanotubes provides a template for the polymer chains to form extended structures, via the π-π interactions and H-bond formation, that could result in a partial covering of nanotube walls with PEDOT-PSS chains. Although the numbers of CNTs are very few with respect to the polymer chains, the larger dimensions of the former and the molecular recognition features could enable the assembly of chains onto the nanotube walls. The schematic in Fig. 4 shows this conformation of PEDOT-PSS around the nanotubes in the nanocomposite, in agreement with the PDF analysis.

Since the electronic properties of polymer nanocomposites are sensitive to the sample preparation that controls the chain conformation and film morphology, it is important to understand how these structural features affect the charge transport in these systems. The temperature dependence of normalised resistivity [$\rho$ (T) / $\rho$ (300 K)] vs. temperature (4-300 K) of 3 % CNT-PEDOT-PSS film is compared with the pristine PEDOT-PSS and also CNT pellet, as shown in Fig. 5. Inset shows the resistivity ratio for CNT pellet for a better view. It seems surprising that the behaviour of resistivity in the two PEDOT-PSS systems is nearly identical, although the presence of CNT is expected to enhance the charge transport in the nanocomposite. The SAXS data have helped to resolve this issue; as the PEDOT-PSS globules adhered onto the walls of nanotubes decrease the inter-tube interactions, the role of CNT in the net charge transport is weakened considerably. However, the data for T < 4 K, as shown in Fig. 6, show the subtle role of CNTs at very low temperature charge transport. Furthermore, pristine

PEDOT-PSS shows very strong temperature dependence of resistivity, unlike other conducting polymers, due to its special structure [short segments of PEDOT attached to long chain PSS]; and it goes to a deep insulating state at T < 10 K. This indicates the role of small polarons in charge transport, since it is known that low mobility of small polarons requires large thermal activation for hopping. However, the presence of CNTs in the composites slows down this rapid increase of resistivity at T < 10 K. Resistivity ratio for CNT pellets in the inset of Fig. 5 shows that the resistivity increases by a factor of two at low temperatures, mainly due to the weak inter-tube transport; yet the large finite conductivity (1000 S/cm) at T ~1 K shows the expected intrinsic metallic nature of these MWNTs.

The resistivity ($\rho$) vs. temperature (T) plot for for 0.03, 0.3 and 3 wt % CNT-PEDOT-PSS nanocomposites are shown in Fig. 6. Apparently, considerable variation in the CNT content does not affect the behaviour of resistivity, and the samples show very similar characteristics down to 1.4 K. As the temperature decreases from 300 to 4 K, the resistivity increases by four orders of magnitude. Although 3 wt % is quite adequate for observing the percolation threshold in usual CNT composites, the weak inter-tube transport due to the presence of PEDOT-PSS globules, as inferred from the SAXS data, makes the transport via the vast polymer matrix more dominant. As mentioned before in case of the pristine polymer, the strongly activated small polaron hopping mechanism remains significant even in presence of CNTs at T > 4 K.

However, at T < 4 K, an anomalous decrease in resistivity has been observed in all the three samples. This intriguing behaviour is found to be reproducible in several samples containing CNTs, and also with CNTs obtained from different sources. In such highly resistive samples this type of behaviour is quite rare, and a possibility of some contribution from the usual Joule heating at T < 4 K has to be taken into account. Nevertheless this heating aspect cannot justify for an order of magnitude drop in resistivity that is observed in this case, since the value of currents used in the measurements are typically less than 1 µA and the samples are immersed in liquid helium during the measurements. This intriguing behaviour has been investigated further as a function of both current and magnetic field, as shown in the insets of Fig. 6; and both show corresponding effect on this variation in resistivity. It is interesting to note that the magnetic field and current dependence of this transition is rather unique. The temperature dependence becomes stronger with increasing magnetic field unlike usual superconducting transition where the transition is weakened by introduction of magnetic field. This indicates the presence of CNTs in PEDOT-PSS matrix favours the possibility of another mechanism of transport to trigger at T < 4 K.

In case of small polaron transport, Bryskin *et al* [20] has proposed the possibility for polarons to undergo intersite tunnel percolation at low temperatures, as a result the temperature coefficient of resistivity (TCR) can vary considerably. The usual phonon-assisted hopping mechanism of small polarons, at higher temperatures, gives rise to the insulating behaviour, and the unusual tunnel transport of polarons at lower temperatures can yield a positive TCR. Such a mixing of both hopping and tunnel transport can occur in disordered systems with strong electron-phonon coupling, in which the charge transport is due to both classical jumps and quantum mechanical intersite tunnelling. The latter contribution to conductivity is usually very small due to the random distribution of polaron energy at localized sites. During tunnel transition between any two sites, a single strong scatterer is chosen from the whole set of scatterers, and it is possible if the energy difference is less than $k_BT$; and the phase coherence of the tunnelling polaron is preserved [20]. Also due to strong electron-phonon coupling and broadening of energy levels, the tunnel transport can be facilitated since the energy difference between the sites is lowered. Nevertheless this process becomes difficult when the extent of disorder is too large, as in case of most of the conducting polymers. Since individual CNTs are highly structured while compared to polymer chains, the presence of CNTs in PEDOT-PSS matrix favours the enhancement of tunnel transport at low temperatures. In such case, the hopping and tunnelling contributions to the total conductivity become comparable at some temperature; below which non-activated tunnel transport dominates and above which activated hopping dominates. This could also explain why magnetic field and electric field (current) could tune the maxima in resistivity, by varying the contributions from both hopping and tunnelling, as shown in the insets of Fig. 6. Moreover, the presence of metallic CNTs could be the reason why the tunnel transport shows a temperature dependence. This model could give a satisfactory explanation for this large drop in resistivity at T < 4 K; however, more theoretical investigations would facilitate to sort out the parametric details.

For a quantitative analysis of the resistivity data for T > 4 K, the reduced activation energy (W) is estimated as [21,22] :

$$W(T) = -\partial \ln R(T) / \partial \ln T \qquad (3)$$

The W-plot, as shown in the inset of Fig. 7, yields a slope of -0.5 indicating that the system is in insulating regime in the temperature range 4 – 300 K with the resisitvity following a stretched exponential dependence (exponent ~ 0.5). Fig. 7 shows the logarithmic resistance plotted vs. $T^{-1/2}$ for the 3% CNT-PEDOT-PSS sample. The $T^{-1/2}$ dependence of lnR can arise from various transport mechanisms like 1D-VRH, Efrös-Shklovskii (ES)-VRH and transport in granular metal [21,23,24]. Among these the ES-VRH can be ruled out, since the Coulomb gap is supposed to continually open up at low temperatures (typically T < 20 K), and the resistance rapidly increases at lower temperatures, which is not the case here. In granular metallic systems the $T^{-1/2}$ fit is supposed to extend to higher temperatures (typically 1 K< T <100 K) as the charging energies are typically around these values. The $T^{-1/2}$ fit for these samples, in the range of 4 K < T < 50 K, suggests that the 1-D hopping of small polarons along the PEDOT segments attached to the PSS template is the plausible mechanism; and also explains why the temperature dependence of resistivity is not sensitive to the variation of CNT content, in this range of temperature. This model based on small polaron hopping and tunnelling could give an interpretation for both the strong temperature dependence of resistivity at T > 4 K and the drop in resistivity at T < 4 K. The temperature dependence of resistivity for 1-D Mott- VRH in disordered systems is given by:

$$R(T) = R_0 \exp[(T_0/T)^{1/2}] \qquad (4)$$

where $T_o = 24/\pi k_B L_c^3 N(E_F)$ is the characteristic Mott temperature, $L_c$ the localization length and $N(E_F)$ is the density of states at the Fermi level. The fit to Eq. 4 in Fig. 7 yield the values of $R_0$ and $T_0$ as 20 Ω and 784 K respectively.

The low temperature resistivity data in CNT-PEDOT-PSS nanocomposites is further investigated with the help of magnetoresistance (MR) measurements as a local probe, as shown in Fig. 8 (a). The MR data for CNT pressed pellet is shown in Fig. 8 (b) for comparison. It is known that the conductivity and MR in CNT can vary widely due to the contributions from defects, packing and alignment of nanotubes. However the observation of negative MR ~ 9 % in Fig. 8 (b) indicates that the intrinsic metallic nature of CNTs is not adversely affected due to disorder. Both conductivity and MR in CNT bundles have been explained using the weak localization (WL) and electron-electron interaction (EEI) model [25]. The present data, thus, agrees with existing earlier reports; the strong negative MR at low fields is due to the WL contribution, and its tendency to saturate at higher fields is due to the superimposing contribution of positive MR due to EEI.

Surprisingly for 3 % CNT-PEDOT-PSS nanocomposite as in Fig. 8 (a), the MR data is positive at T > 4 K, and it is negative at T < 4 K, which is reproducible in several samples. This type of MR is hardly observed in conventional hopping systems. The positive MR at 4.2 K and 11 T is ~ 30 % which is quite low, unlike the large positive MR observed in usual hopping systems. This is due to the presence of CNTs that contributes in a resulting lowered magnitude of MR.  The low value of positive MR also indicates that the response of low mobility polarons to the magnetic field is rather weak, as is expected. However, the unusually large negative MR at T < 4K especially at 1.3 K (~ 80 %) is quite unprecedented in these types of systems. The earlier MR studies in several conducting polymers like polypyrrole, polyaniline, PEDOT, etc., and also in MWNT-insulating polymer composites, have not shown any comparable results. Typically the systems close to metal-insulator transition show a small positive MR, insulating ones have a large positive MR and metallic samples have a mix of small positive and negative due to the contributions from weak localization and electron-electron interactions [26,27].

The MR data above 4 K can be quantitatively analyzed based on the 1D-VRH model prescribed for the resistivity data for the temperature range 4 – 300 K. In the strongly localized regime, presence of an external magnetic field leads to shrinkage of the overlap of wavefunctions of the charge carriers. As a result the probability of hop between two sites is reduced and a large positive magnetoresistance (MR) is observed [28,29].

For 1D-VRH transport, the low-field positive MR is governed by the equation:

$$\ln[\rho(H)/\rho(0)] = t(L_c/\eta)^4 (T_0/T)^y \qquad (5)$$

where $\eta = \sqrt{\hbar/eH}$ is the magnetic length and $L_c$ is the localization length, t = 0.0015, and y = 3/2 [6].

Fig. 9 shows ln [ρ (H) / ρ (0)] vs. $H^2$ plots at 4 K and 10 K for the positive MR data of 3 % CNT-PEDOT-PSS nanocomposite. Using the value of $T_0$ from the resistivity data, the values of $L_c$ obtained from the fits to Eq. 5 are 4.5 nm and 3 nm at 4 K and 10 K respectively. The values lie between the small localization lengths reported for conducting polymers (~ 1–2 nm) and comparatively larger value for MWNTs (~ 10 nm) [6,30]. It is thus evident that the localization of the polymer matrix is enhanced due to the presence of CNTs indicating an overlap of wavefunctions between carriers of polymer and MWNTs especially at lower temperatures.

Although VRH transport is usually observed in disorderly materials like composites, as the inhomogeneity of the system increases, fluctuation induced tunnelling (FIT) model is used to fit the data especially at low temperatures, for conducting nanofillers in an insulating matrix [31]. Here the tunnelling is characterised by charge transport across insulating barriers in the conducting pathways between conducting regions. Thermally activated voltage fluctuations across the junctions can lead to temperature independent conductivity at very low temperatures. Recently a combined FIT and VRH model is used to analyse the data in MWNT-PMMA composite [4]. However, fitting the data to this mutiparameter model often leads to physically unrealistic values and ambiguous results. Furthermore, in CNT-conducting polymer system, as in the present case, the conducting matrix does not favour the voltage fluctuations to occur across the tunnel junctions. Nevertheless, if the proposed model of the tunnel transport of small polarons [20] at T < 4 K is viable, then the magnetic field could lower the tunnel barriers, especially at the polymer-CNT interface, as a result a large negative MR is possible, since the tunneled polarons are expected to have a higher mobility via the CNTs. This type of field-induced delocalization, especially in low dimensional systems has been investigated theoretically; and the magnetic field is shown to enhance the interchain hopping integral, as a result a large negative MR is possible [32]. In such case, the theoretical model [32] proposes this screnario: $\tau_{tr} < \tau_H < \tau_{ph}$, which could produce a large negative MR with a $H^2$ dependence at low fields and saturates at H > 4 T; $\tau_{tr}$, $\tau_H$ and $\tau_{ph}$ being longitudinal transport time, phase breaking time due to magnetic field and phase breaking time due to phonon forward scattering respectively.

## 4. Conclusions

Conformation and charge conduction mechanisms in MWNT-PEDOT-PSS nanocomposites are studied. The results show how the structural features present in the system affect the overall charge transport.

The SAXS studies in conjugated polymers and its composites with CNT show that supramolecular scale interactions modify the structural properties. Partial covering of the CNTs by the PEDOT-PSS globules which is evident from SAXS data weakens the inter-tube interactions of CNTs. Since CNTs usually have a tendency to bundle, PEDOT-PSS can assist in separating them out. Interestingly, these solution-state conformational features provide clues for the repercussions in the charge transport properties in solid-state films.

The temperature and magnetic field dependence of resistivity in nanocomposites of MWNT-PEDOT-PSS are analyzed within a consistent framework. Charge transport is mainly governed by small polarons of the conducting polymer as the interaction between nanotubes is sufficiently reduced by the polymer globules adhered to the nanotube walls. At $T > 4$ K, 1-D VRH is dominant, which results in a positive MR. However, the anomalous drop in resistance and the large negative MR at $T < 4$ K are attributed to the tunnel transport of small polarons in this temperature regime. The presence of quasi-1D MWNTs helps in lowering the tunnel barriers thereby enhancing delocalization across the nanotube-polymer boundary that leads to a large negative MR.

FIGURE CAPTIONS

Figure 1. TEM image of 3 wt. % CNT-PEDOT-PSS nanocomposite.

Figure 2. I vs. q profiles for different concentrations of CNTs in aqueous PEDOT-PSS. Baselines are shifted for clarity. The inset shows I vs. q profile of 1 % CNTW suspension. Fits to different q-ranges show the corresponding slopes.

Figure 3. I vs q profile of PEDOT-PSS in nanocomposite obtained by subtraction of CNT-water (CNTW) data from the 1 % CNT-PEDOT-PSS (CNTP) data. Solid lines are fits showing the corresponding slopes of different q-ranges. Inset shows aq. PEDOT-PSS profile. Baselines are shifted for clarity.

Figure 4. Pair distribution function of PEDOT-PSS in nanocomposite compared to that in water. Baselines are shifted for clarity. Schematic shows conformation and organization of the elongated PEDOT-PSS globules on and around CNT.

Figure 5. Resistivity ratio [ρ (T) / [ρ (300 K)] vs. temperature plots for 3 % CNT-PEDOT-PSS as compared to pristine PEDOT-PSS and MWNT pellet. Inset shows the resistivity ratio plot for MWNT pellet for a better view.

Figure 6. Resistivity vs. temperature for 0.03, 0.3 and 3 % CNT-PEDOT-PSS nanocomposites. Top-right inset shows magnetic field dependence and bottom-left inset shows current dependence of ρ vs. T plot for the 3 % nanocomposite.

Figure 7. lnR vs. T-1/2 for the 3 % CNT-PEDOT-PSS; solid line is a fit to Eq. 4. Inset shows reduced activation energy W vs. T with a slope -0.5.

Figure 8. Magnetoresistance vs. magnetic field at different temperatures for (a) 3 % CNT-PEDOT-PSS and (b) MWNT pellets.

Figure 9. ln [ρ (H) / ρ (0)] vs. $H^2$ plots for the positive MR data of 3 % CNT-PEDOT-PSS as shown in Fig. 8 (a). Solid lines are fits to Eq. 5.

Figures

Fig. 1. TEM image of 3 wt. % CNT-PEDOT-PSS nanocomposite

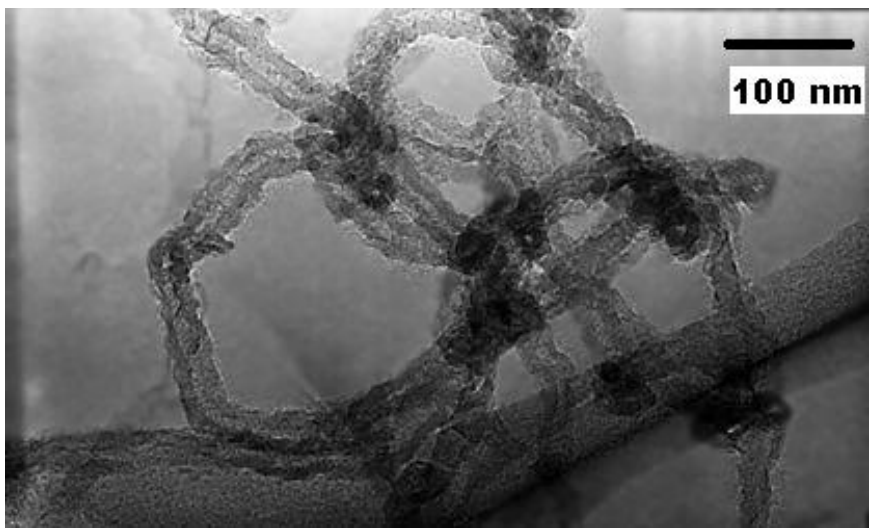

Fig. 2. I vs. q profiles for different concentrations of CNTs in aqueous PEDOT-PSS. Baselines are shifted for clarity. The inset shows I vs. q profile of 1 wt. % CNTW suspension. Fits to different q-ranges show the corresponding slopes.

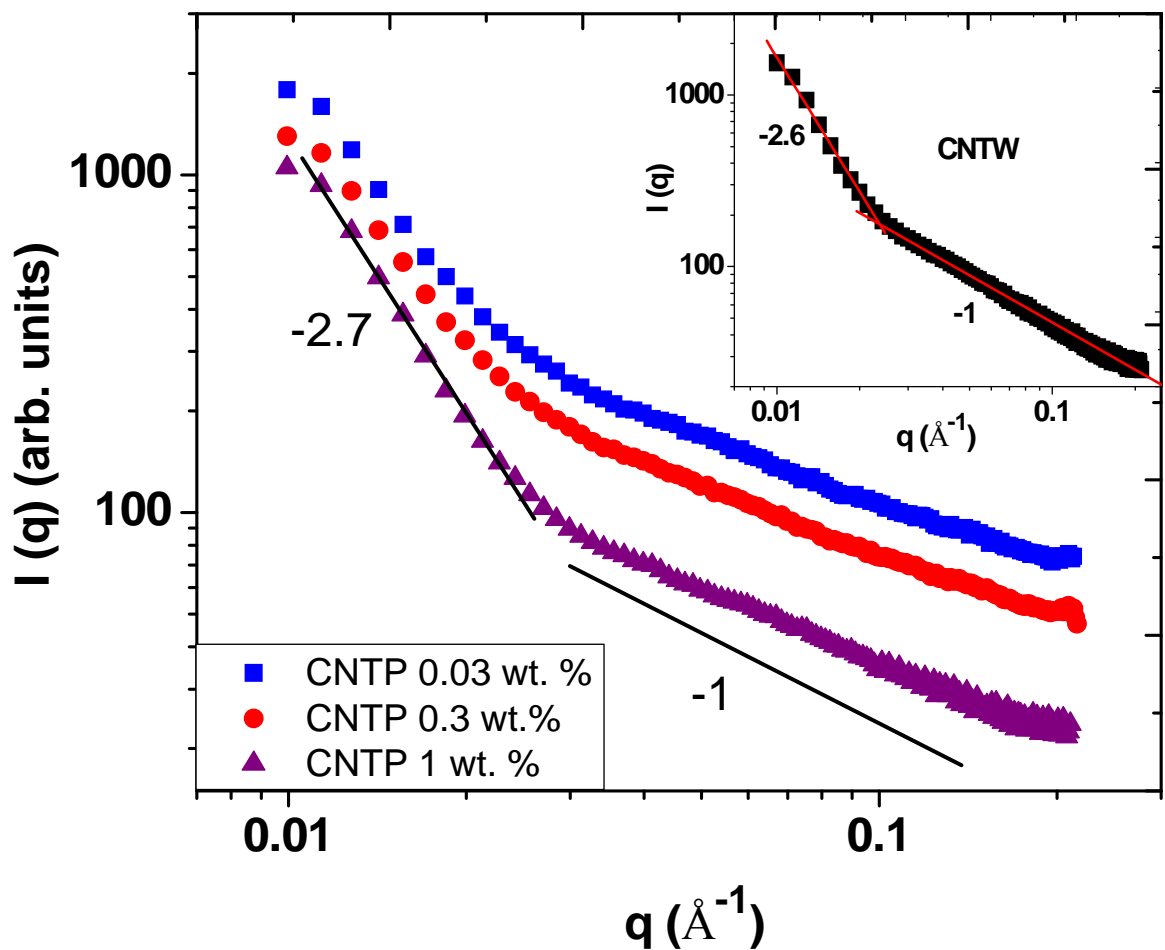

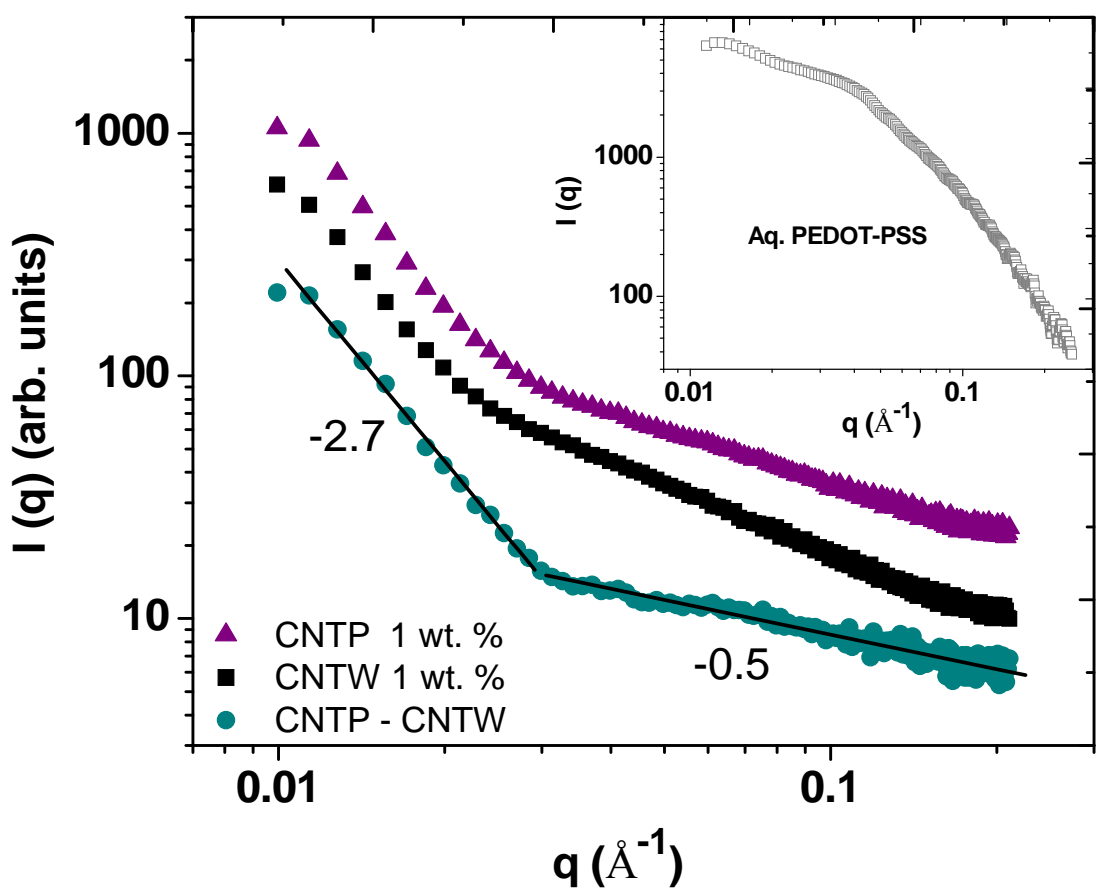

Fig. 3. I vs q profile of PEDOT-PSS in nanocomposite obtained by subtraction of CNT-water (CNTW) data from the 1 wt. % CNT-PEDOT-PSS (CNTP) data. Solid lines are fits showing the corresponding slopes of different q-ranges. Inset shows aq. PEDOT-PSS profile. Baselines are shifted for clarity.

Fig. 4. Pair distribution function of PEDOT-PSS in nanocomposite compared to that in water. Baselines are shifted for clarity. Schematic shows conformation and organization of the elongated PEDOT-PSS globules on and around CNT.

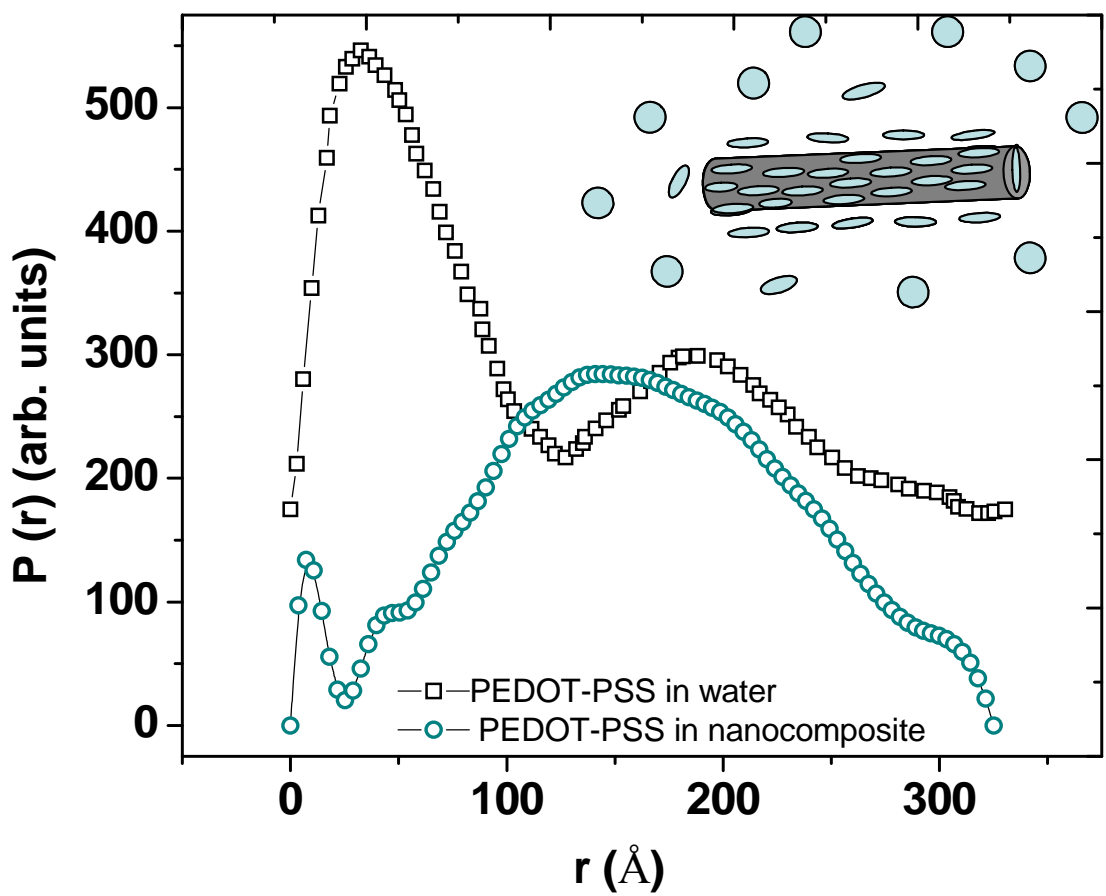

Fig. 5. Resistivity ratio [ρ (T) / ρ (0)] vs. temperature plots for 3 wt. % CNT-PEDOT-PSS as compared to pristine PEDOT-PSS and MWNT pellet. Inset shows the resistivity ratio plot for MWNT pellet for a better view.

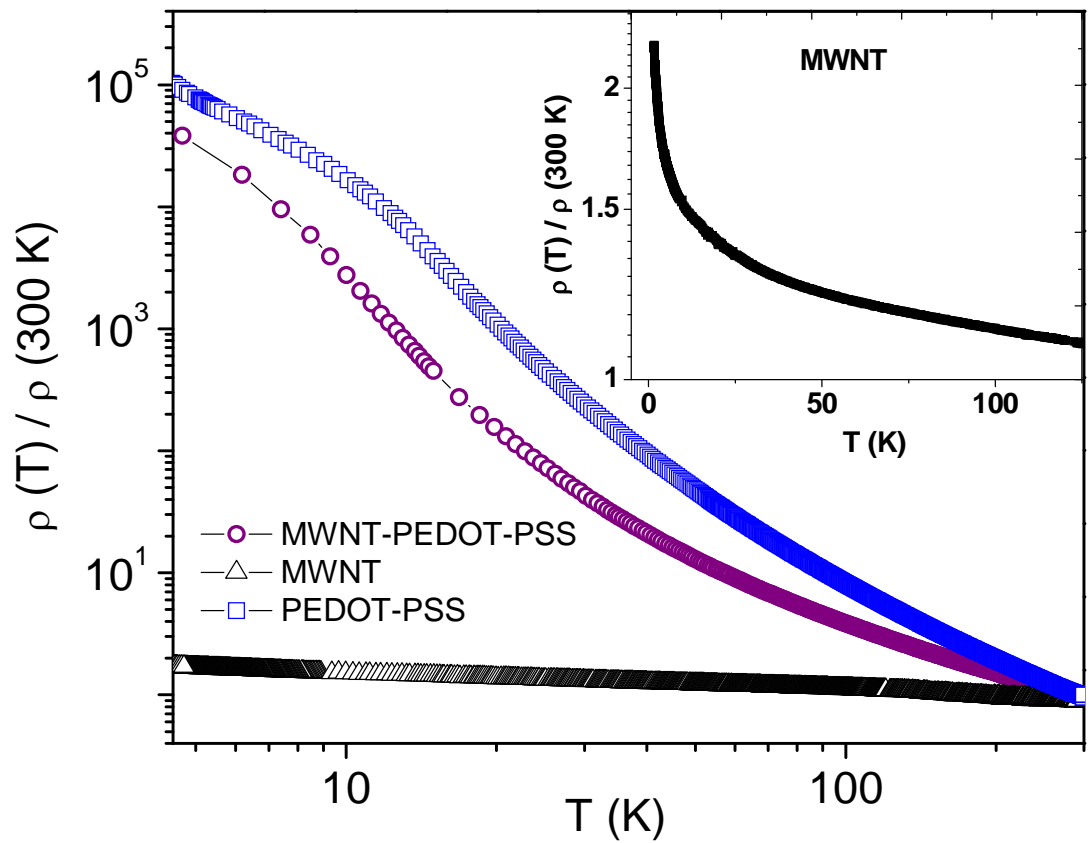

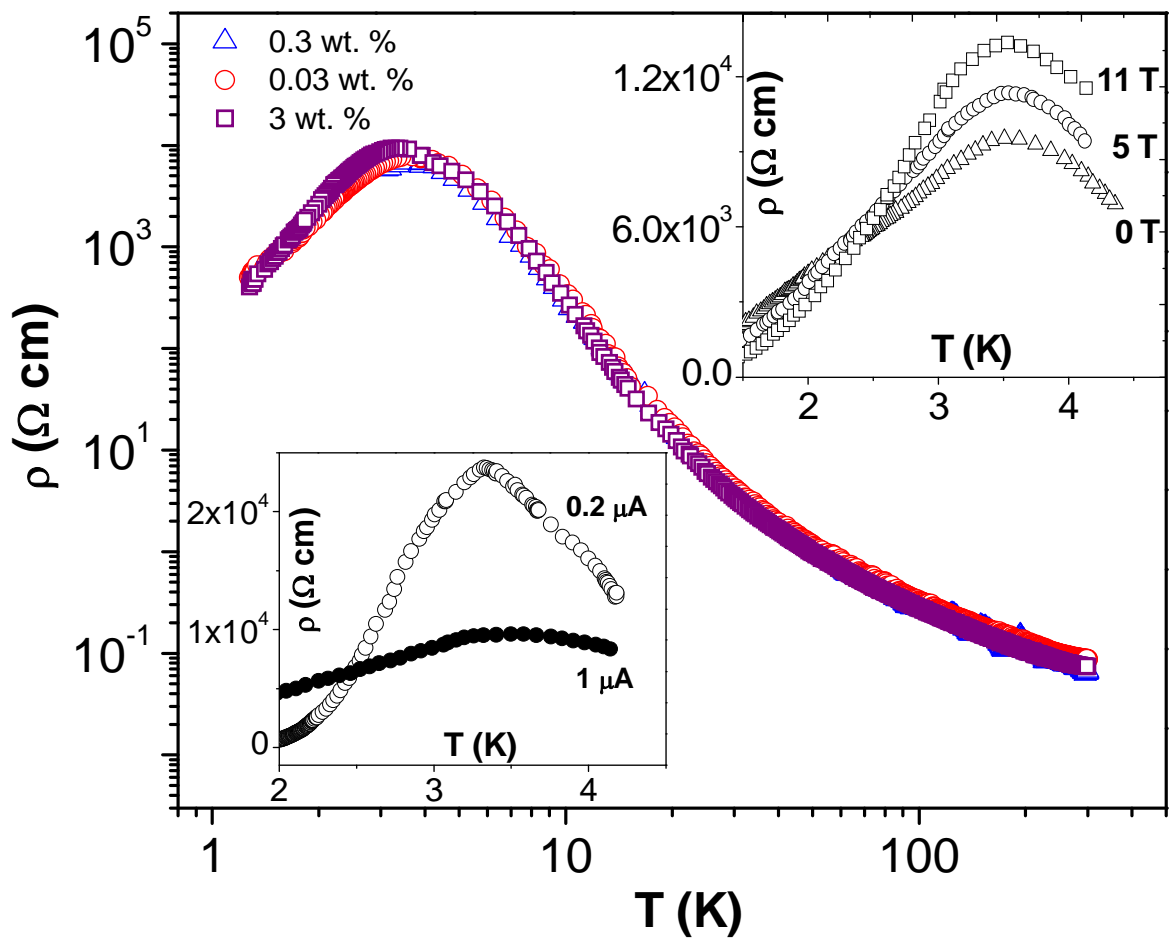

Fig. 6. Resistivity vs. temperature for 0.03, 0.3 and 3 wt. % CNT-PEDOT-PSS nanocomposites. Top-right inset shows magnetic field dependence and bottom-left inset shows current dependence of ρ vs. T plot for the 3 wt. % nanocomposite.

Fig. 7. lnR vs. $T^{-1/2}$ for the 3 wt. % CNT-PEDOT-PSS; solid line is a fit to Eq. 4. Inset shows reduced activation energy W vs. T with a slope -0.5.

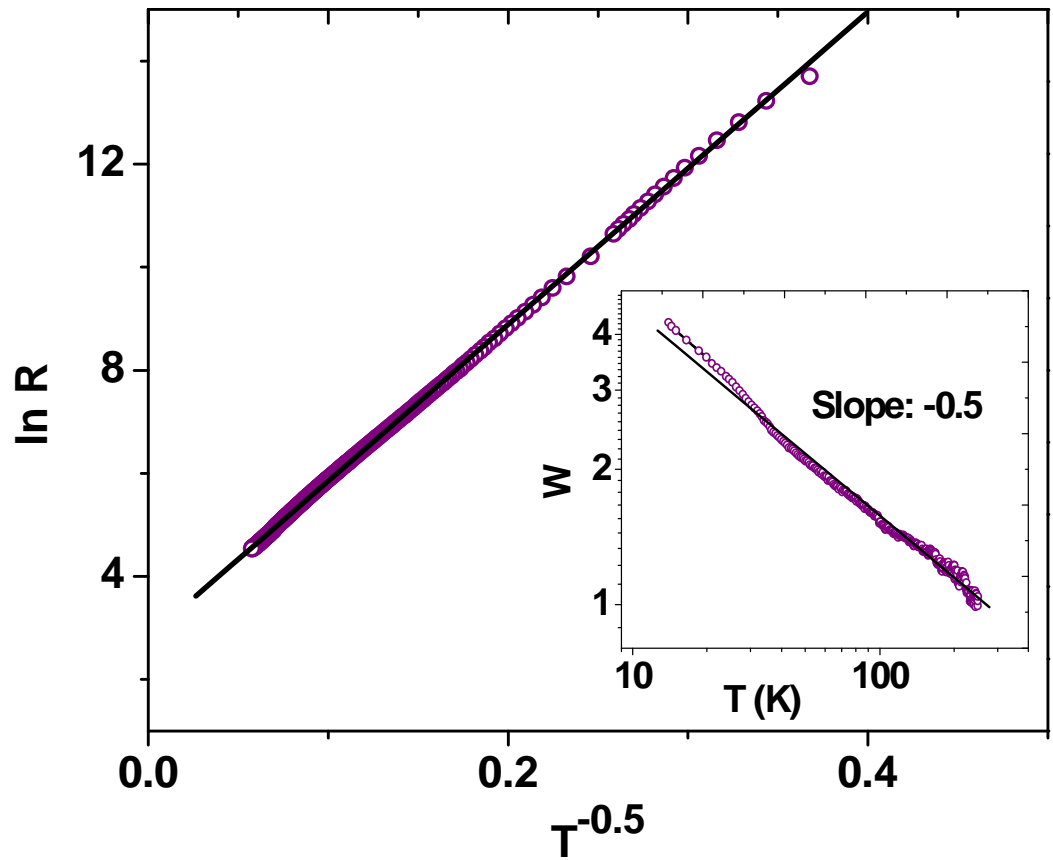

Fig. 8. Magnetoresistance vs. magnetic field at different temperatures for (a) 3 wt. % CNT-PEDOT-PSS and (b) MWNT pellets.

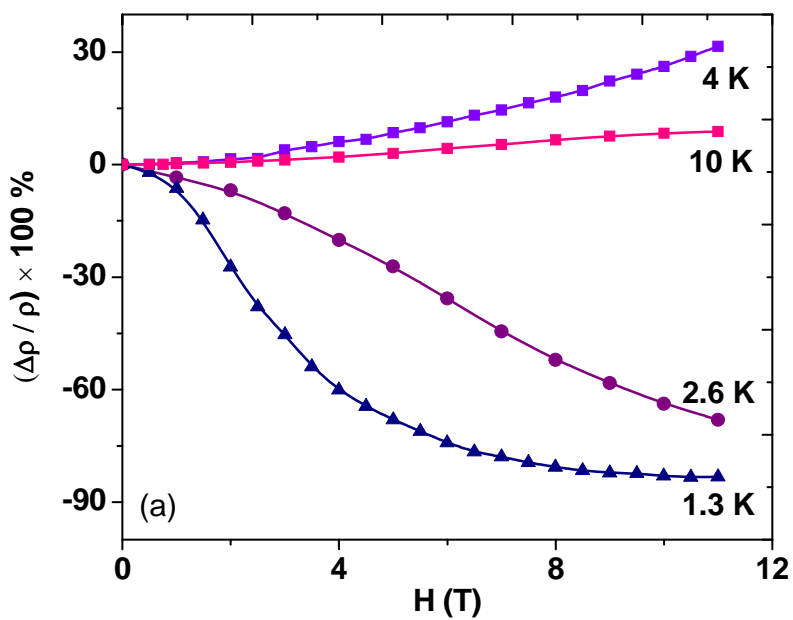

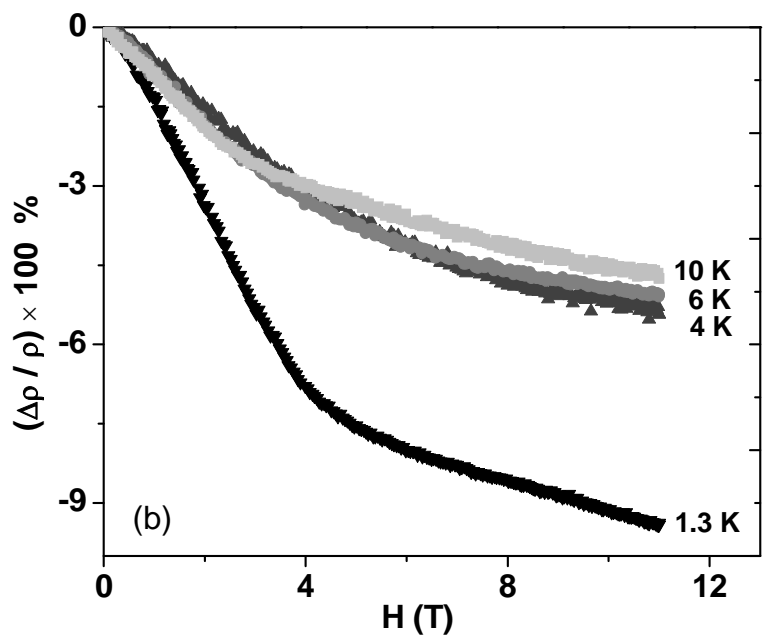

Fig. 9. ln [ρ (H) / ρ (0)] vs. H² plots for the positive MR data of 3 wt. % CNT-PEDOT-PSS as in Fig. 8 (a). Solid lines are fits to Eq. 5.

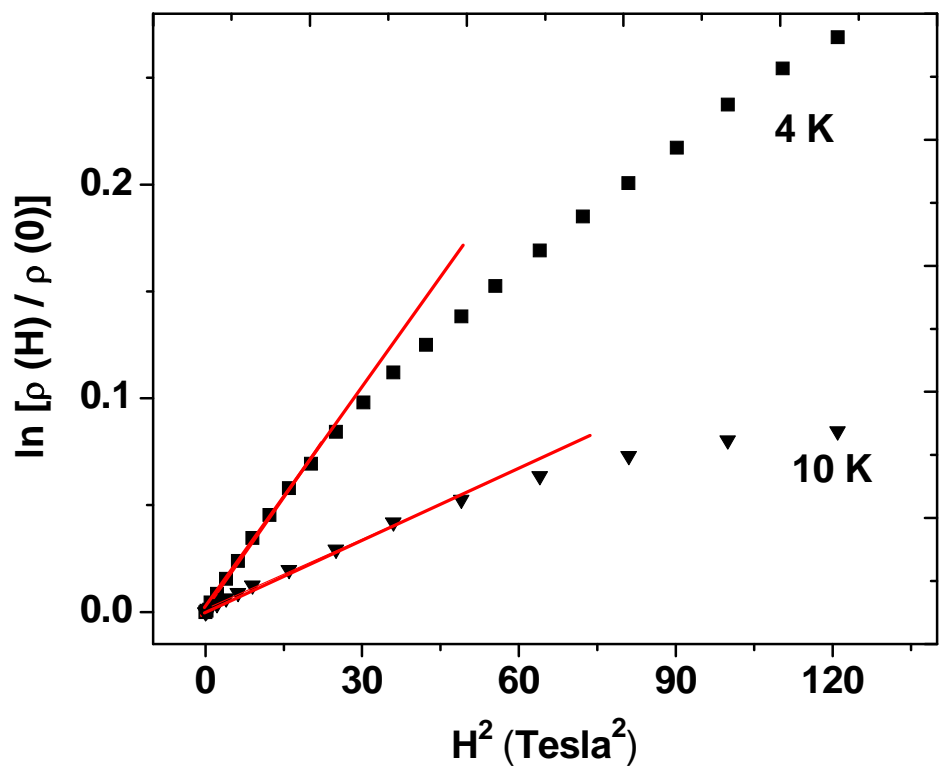